\documentclass[12]{article}
\usepackage{amsmath}
\usepackage{amsfonts}
\usepackage{amssymb}
\usepackage{amsthm}
\usepackage{color}

\newcommand{\R}{{\rm I}\kern-0.18em{\rm R}}
\newcommand{\1}{{\rm 1}\kern-0.25em{\rm I}}
\newcommand{\E}{{\rm I}\kern-0.18em{\rm E}}
\newcommand{\p}{{\rm I}\kern-0.18em{\rm P}}


\usepackage{epsfig}

\usepackage{fancyhdr}

\usepackage{graphics}

\usepackage{amsopn}  %%% for new Math operators
\usepackage{amssymb}
\usepackage{amsxtra}
\usepackage{amsfonts}

\usepackage{afterpage}
\usepackage{eucal}
\usepackage{eufrak}

\usepackage{enumerate}

\allowbreak

\def\fnote#1{\footnote}

\newcommand{\bea}{\begin{eqnarray}}

\newcommand{\eea}{\end{eqnarray}}

\newcommand{\beas}{\begin{eqnarray*}}
\newcommand{\eeas}{\end{eqnarray*}}

\author{G Temnov\footnote{Department of Probability and Statistics, MFF, Charles University, Prague-8, 18675, Czech Republic}}
\title{Analysis of Ornstein-Uhlenbeck process stopped at maximum drawdown 
and application to trading strategies with trailing stops }
\date{}

\begin{document}
\maketitle

\begin{abstract}
We propose a strategy for automated trading, outline theoretical justification of 
the profitability of this strategy and overview the hypothetical results in application 
to currency pairs trading.

The proposed methodology relies on the assumption that processes reflecting the dynamics of currency exchange rates are in a certain sense similar to the class of Ornstein-Uhlenbeck processes and exhibits the mean reverting property.

In order to describe the quantitative characteristics of the projected return of the strategy,
we derive the explicit expression for the running maximum of the Ornstein-Uhlenbeck
(OU) process stopped at maximum drawdown and look at the correspondence
between derived characteristics and the observed ones.
\end{abstract}

{\bf keywords:} automated trading strategy, Ornstein-Uhlenbeck process

\section{Structure of the work and description of the general idea}

This paper is structured as follows.

The current chapter outlines the general idea and the setting of the strategy as well as the motivation behind
the underlying research.

Chapter 2 describes the direct intuitive scheme for the optimization of the key parameters of the strategy from historical data, quotes the results of the strategy's returns and poses questions for the further analysis.

Chapter 3 highlights the main analytical point of this work - explicit formula for the distribution of the running maximum of the OU process stopped at maximum drawdown, and discusses how it relates to
the strategy and which characteristics of the strategy's returns can be derived from the running maximum distribution.

Chapter 4 overviews the parameters' estimation methods for OU process and refers to the results
of the estimation on the basis of the historical data that we use for testing the strategy's efficiency.

Chapter 5 summarizes the correspondence between the actual results and analytical estimates, and provides the outlook on the further optimization.

%Chapter 4 quotes the results of implementing the strategy using %different methods of parameters'
%adjustment and updating.

\vspace*{0.8cm}

{\bf General idea of the strategy} \par

As mentioned, the proposed methodology relies on the assumption that processes reflecting the dynamics of currency exchange rates are in a certain sense similar to the class of OU processes and exhibit the mean reverting property.

In other words, as such process deviates from its current mean, a certain "force" tends to revert it back to its mean value. This property can be exploited in the context of ForEx markets dynamics: Due to occasional sparks of increased volatility usually caused by economic factors, the exchange rate may burst out creating a potential force that tends to drive the process back to its mean trend level. Opening a long or short position, contrary to the direction of the outburst, may allow to take advantage of this driving force.

\vspace*{0.3cm}
Observations of the weekly EUR/USD dynamics confirm that the profile indicating the configuration "outburst followed by a movement in the opposite direction" (in technical terms usually interpreted as correction or consolidation) is frequently observed and agrees with the above described intuition.

The strategy that relies on the above observation can loosely be outlined as follows:

\begin{itemize}
\item
At the start of each week of trading, set up an exchange rate level that can serve as "zero-level" (usually, weekend rate or the opening of the week rate) and pre-set the triggers for position opening (described next).

\item
The position will open if the rate rises above a predetermined level U (short position opening) or drops below the level D (long position opening), where U and D are measured from the "zero-level'. The position opens automatically depending on which of the levels, $U$ or $D$, is hit first.

\item
As soon as the position is opened, the trailing stop (TS) and the profit call (CP) levels are affiliated with the position so that it will close automatically as soon as either of these stops work out (or it will close at the weekly trading closure if none of them is hit).
\end{itemize}

To summarize, the above strategy is simply designed to take advantage of the correction that often follows the initial outburst, usually near the start of the trading week -- and therefore use the driving force that reverts the process back to its long-term mean, in favour of the trader.

\vspace*{0.2cm}
Of course, it is possible that the initial price movement that triggers the opening of the position is actually a reflection of the trend rather than an "outburst", so that the opened position would in fact be held against the trend and would therefore be a potential loss.

However, long history of observations on EUR/USD trading pair show that configurations with drawdown/drawup following the outburst up/down are observed within almost each of the trading weeks, while the pure-trend configuration is a relatively rare situation.

\vspace*{0.3cm}
The above argument is nevertheless purely intuitive and we will, of course, need a more solid probabilistic and statistical reasoning to justify the potential profitability of the strategy. We address the probability context of the model in the following chapters, where we will consider the analytical representation of the distribution of the return of the strategy as well as the estimation of parameters of the underlying process, under the hypothesis that this process is of an OU type.

\vspace*{0.3cm}
Taking another look at the idea of the strategy and its implementation, we note that the key problem in this context is the task of optimizing the parameters U, D, TS, CP. We will be looking for such a set (U, D, TS, CP) that would produce the maximum aggregated profit for the strategy.
% – in other words, summing up the pip differences in all the closed positions.

\vspace*{0.8cm}

\section{Practical realization of the strategy and empirical scheme for parameters optimization}

Before proceeding to analytical study in the following chapters, we start with a straightforward
scheme of the strategy implementation, as the practical results are likely to highlight the
strategy's potential and indicate the points of the further analysis.

At first instance, we estimate the parameters in a simplistic way, via the "ad hoc" rule: assuming that the 
 historical data that is constantly revealed, can be used for optimization of the parameters $(U, D, TS, PC)$
directly by maximizing the "what if" returns of the strategy over the past years, the scheme can
roughly be described as:

\begin{itemize}
\item At the start of year $N$, use the available data up to and including year $(N-1)$ to
estimate the set $(U, D, TS, PC)$ of parameters such that the aggregate return over the past period
(say, over year $(N-1)$ only if no earlier data is available) would be largest possible.

\vspace*{-0.15cm}
\item
Implementing the strategy during the year $N$, use the set $(U, D, TS, PC)$ of parameters estimated at the previous step.

\vspace*{-0.15cm}
\item
By the start of year $N+1$, use the actual result of the strategy's return over the year $N$, and
in addition use the data of the year $N$ that is now available along with previous history, to simulate possible returns with other choices of parameters and therefore to perform the  "what if" analysis and obtain the updated estimate of the set $(U, D, TS, PC)$ of parameters such that the aggregate return over the year $N$ would be largest possible (or, alternatively, the aggregate return over year $(N-1)$ {\bf and} year $N$ together would have been largest).

\vspace*{-0.15cm}
\item
At the start of each following year $N+i$ ($i>1$), repeat the above updating scheme, using different 
selective data sets (using data of previous year, $(N+i-1)$, only, or using two or more years of history 
that has been revealed by the year $N+i$).
\end{itemize}

In relation to this scheme, note that the parameters' updating can, of course, be performed on a more frequent than yearly basis.

At first instance of the strategy implementation, the optimization of the parameters' set was performed approximately (without using automated optimization techniques which would be appropriate is this case
but would also appear computationally expensive considering the set of $4$, possibly cross-dependent,
parameters). This approximate scheme allowed to draft first conclusions about preferable ways of
parameters' updating.

Figure 1 indicates the histogram of the weekly returns as a result of  the implementation of the strategy with parameters' optimization of the basis of $1$ recent year data only, using the historical data of eur/usd over $4.5$ years from $2011$ to the middle of $2015$.
\vspace*{-0.8cm}
\begin{figure}[ht!]
\centering
\includegraphics[width=70mm]{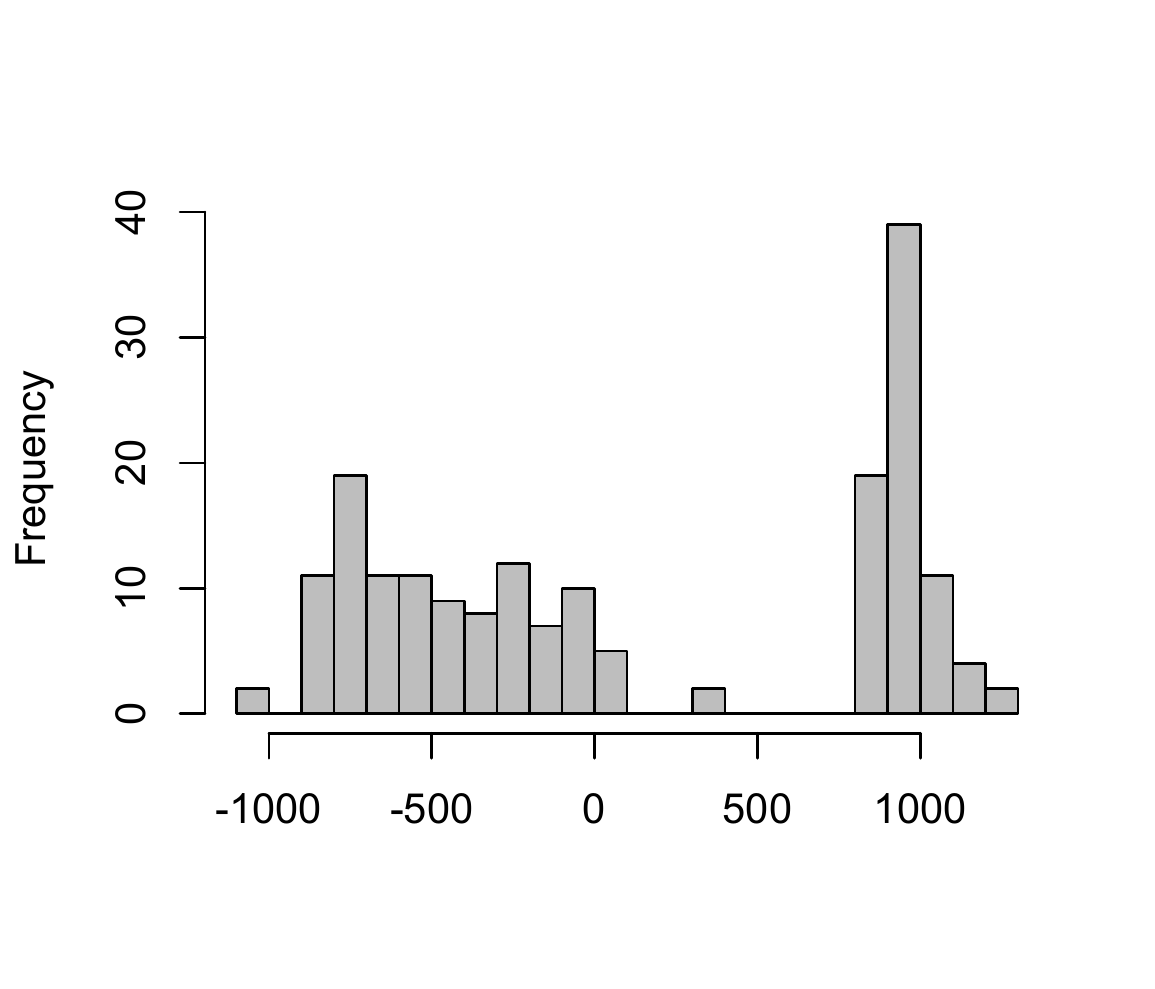}
\vspace*{-1.2cm}
\caption{Weekly returns (in euros) of the test run of the strategy (the actual period covered is from start of $2012$ to the mid of $2015$ (around $182$ weeks)) \label{F1}}
\end{figure}

\vspace*{0.25cm}
We quote the results of different scheme's implementation in Table 1 below, which is structured as follows:
the lines with the year number followed by "e", such as $2011$e etc, indicate the results of 
{\it estimation}, so that the corresponding year's data were used to optimize the parameters to obtain
the largest possible value of the mean weekly returns $\mu$.

The lines with the year number followed by "{\bf A}", on the other hand, indicate the "actual" results,
so that if the parameters estimated from previous years were used during the current year, it would result in the {\it actual}
mean weekly returns value as indicated.

\vspace*{0.15cm}
Note that weekly mean, $\mu$, is expressed in euros.
It is calculated under the assumption that each single position is $1000$ euro with leverage $1:200$, and
the commission for holding the asset overnight is $0.14\%$.

\vspace*{0.15cm}
\vspace{0.1cm} \vspace{0.1cm}\hspace*{-0.4cm}
\begin{tabular}{|p{0.8cm}|p{2.8cm}|p{2.8cm}|p{2.8cm}|p{1.05cm}|} \hline
& \multicolumn{4}{|c|}{ $(U, D, TS, PC),\,\,$ {\bf weekly mean}\,\,$\mu$}  \\ \hline
 & previous year only & {\bf $2$ recent years} & {\bf $3$ recent years} & {\bf $4$ yrs} \\ 
\hline \color{blue}{$2011$e} &
 \color{blue}{ {\small $(19, 20, 51, 58),\mathbf{171}$}  }&
 $\,$ & $\,$ & $\,$ \\
\hline
$2012${\bf A} &
  {\small $(19, 20, 51, 58),\mathbf{122}$} &
 $\,$ & $\,$ & $\,$ \\
\hline
\color{blue}{$2012$e} &
 \color{blue}{ {\small $(45, 33, 52, 61),\mathbf{307}$} }&
  \color{blue}{ {\small $(45, 34, 52, 61),\mathbf{192}$} } & $\,$ & $\,$    \\ 
\hline
$2013${\bf A} &  {\small $(45, 33, 52, 61),\mathbf{224}$} & 
 {\small $(45, 34, 52, 61),\mathbf{199}$} &
  $\,$ & $\,$ \\
\hline
\color{blue}{$2013$e} &  \color{blue}{{\small $(39, 32, 52, 61),\mathbf{251}$}} & 
\color{blue}{ {\small $(44, 33, 52, 61),\mathbf{263}$} }&
 \color{blue}{ {\small $(44, 33, 52, 61),\mathbf{198}$}} & $\,$ \\
 \hline
$2014${\bf A} &  {\small $(39, 32, 52, 61),\mathbf{31}$} &  {\small $(44, 33, 52, 61),\mathbf{5}$} &
  {\small $(44, 33, 52, 61),\mathbf{5}$} &  $\,$  \\
 \hline
\color{blue}{$2014$e} & \color{blue}{ {\small $(20, 22, 51, 59),\mathbf{157}$} } & 
\color{blue}{ {\small $(39, 32, 52, 61),\mathbf{144}$} } &
\color{blue}{ {\small $(45, 33, 52, 61),\mathbf{184}$} } &  tbf  \\
 \hline
$2015${\bf A} &  {\small $(20, 22, 51, 59),\mathbf{128}$} & 
{\small $(39, 32, 52, 61),\mathbf{155}$} &
 {\small $(45, 33, 52, 61),\mathbf{81}$} &  tbf  \\  \hline

\color{blue}{$2015$e} & \color{blue}{ {\small $(23, 21, 55, 63),\mathbf{239}$} }&  
\color{blue}{to be filled}  &
 \color{blue}{to be filled} &  tbf  \\  \hline

$\,$ &  $\,$ &  $\,$ & $\,$ &  $\,$  \\  \hline
\end{tabular}

\vspace*{0.03cm}
Table 1. Results of different schemes of parameters' updating.

\vspace*{0.3cm}
The $P\&L$ process corresponding to these three schemes demonstrate closely related results, as 
indicated on Figure 2.

\vspace*{-0.8cm}
\begin{figure}[ht!]
\centering
\includegraphics[width=90mm]{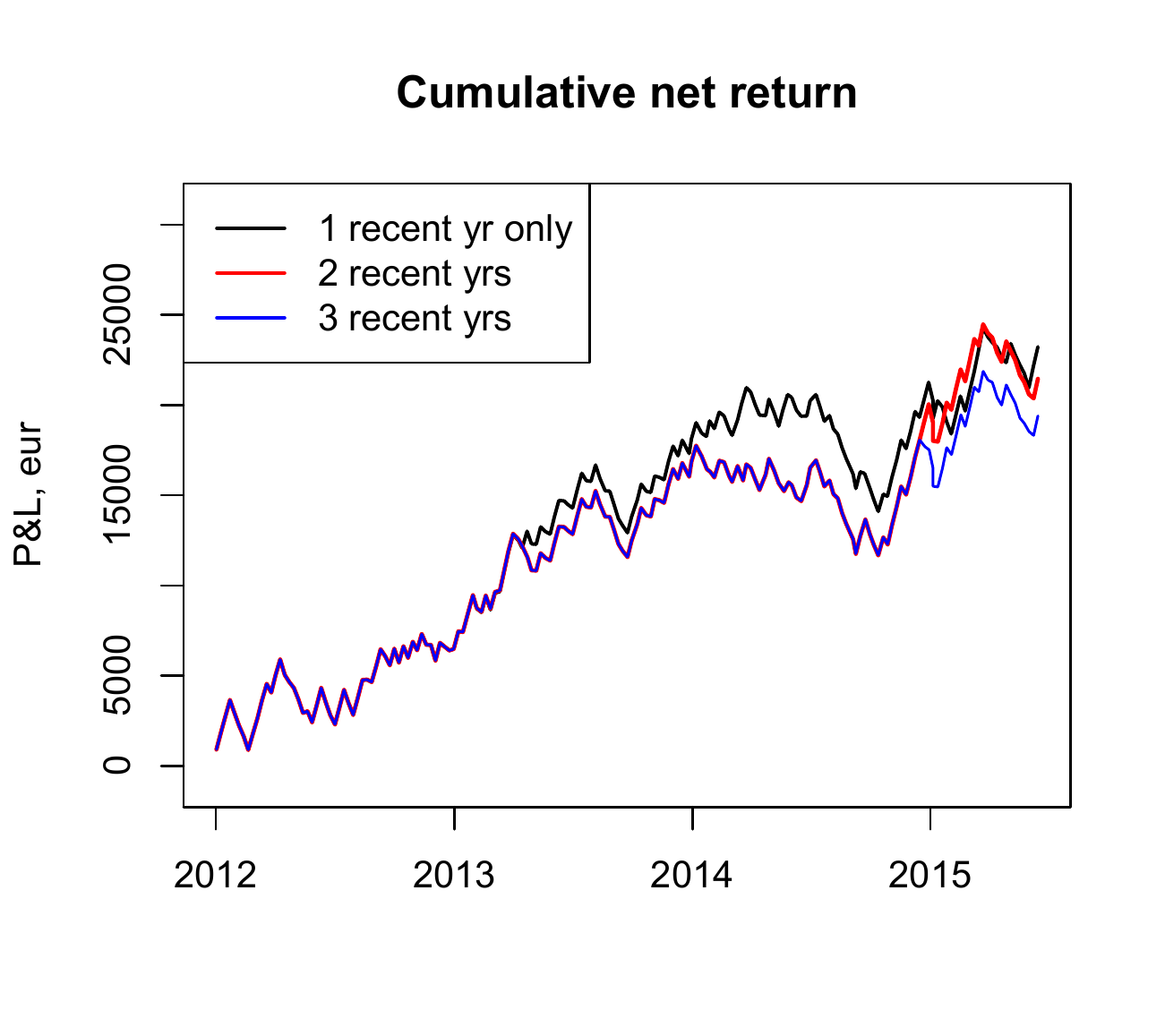}
\vspace*{-1.2cm}
\caption{Dynamics of $P\&L$ according to three schemes of parameters updating:
from the scheme when parameters are estimated from the previous year of
historical data only -- to the one that uses $3$ previous years \label{F1}}
\end{figure}

\vspace*{0.125cm}
As follows from above, the updating scheme that uses
the previous 1-year's history only, appears to lead to the best results.

That might have to do with the "memory" capacity of the underlying diffusion process -- the issue that we will address in the following chapters.

The difference is not too large though, with the following resulting {\it average weekly returns} (in euros):

\vspace*{-0.2cm}
\begin{itemize}
\item $127/$w when using 1 recent year history only.
\vspace*{-0.2cm}
\item $118/$w when using 2 recent years of history.
\vspace*{-0.2cm}
\item $107/$w when using 3 years of history.
\end{itemize}

\vspace*{0.0cm}
As indicated in Figure 2, negative dynamics was observed in the second half of $2014$.
That period corresponds to a fast decrease of eur/usd rate, and the number of weeks
when the rate dropped significantly without being preceded by a considerable drawup, 
lead to the decline of strategy's returns in that period. 

In Figure 3, four weeks taken at random from the second half of $2014$ history, indicate that effect.

In Chapter 4, we will address the problem of how the dynamics of the estimated parameters of the 
underlying OU process react on the time periods of such unusual activity, and in Chapter 5 discuss how the underlying parameter's dynamics can be used to improve the strategy's performance.

\vspace*{-0.3cm}
\begin{figure}[ht!]
\centering
\includegraphics[width=90mm]{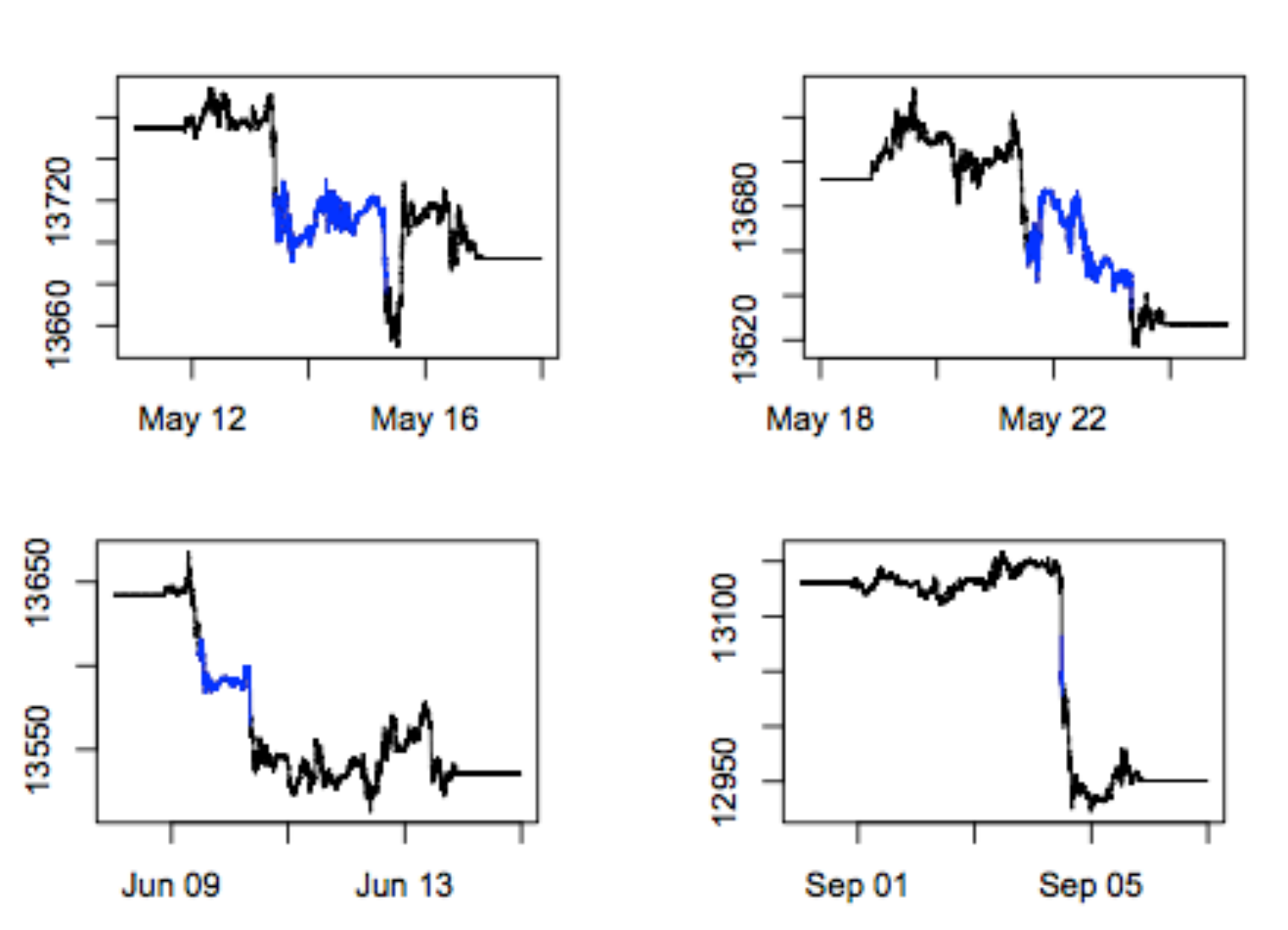}
\vspace*{-0.2cm}
\caption{Four selected trading weeks of EUR/USD from $2014$,
marking the persistent downward trend  \label{F2}}
\end{figure}

%Obviously, such periods are not suitable for..

\vspace*{-0.35cm}

\section{Analytic expression for the maximum of a process stopped at
given level of drawdown}

The general assumption on the dynamics of most financial indices is that the underlying
processes are of diffusion type and can be described by stochastic differential equations (SDE)
of generic form
\[
dX_t = \mu(X_t)dt + \sigma(X_t)dW_t\,,\,\,\,\,X_0=x\,.
\] 

In the context of probabilistic analysis of the proposed strategy and its returns,
the key problem is describing the running maximum of such diffusion process stopped at a given level of drawdown.

Denote the running maximum of process $\{X_t\}$ by $M_t=\sup\limits_{s\in[0,t]}X_s$, and the 
{\it drawdown process} as
\[ DD_t = M_t - X_t\,\,. \]
 
The common way to obtain the running maximum distribution of the process stopped at a fixed drawdown level is by representing this problem as an escape problem (i.e., first passage problem).

Denote by $T_D(a)$ the first passage time of the process $\{DD_t\}$ through the level $a$ \,($a>0$):
\[ T_D(a) = \inf\{t\geqslant 0;\,\, DD_t=a\}\,. \]

The distribution function (cdf) of the r.v. $M_{T_D(a)}$, e.g. the maximum of the process $X_t$ stopped at
the drawdown level $a$, given the initial starting point $x$, is
\begin{equation}\label{runmax}
\mathbb{P}_x\left[ M_{T_D(a)}\leqslant v \right] = 1-e^{-\int\limits_x^v\frac{\Psi(x,z)}
{\int\limits_{z-a}^z\Psi(x,y)dy}dz}\,,
\end{equation}
where the function $\Psi$ is defined as
\begin{equation}\label{Psi}
\Psi(u,z)=e^{-2\int\limits_u^z\gamma(y)dy}\,,\,\,\,\,\mbox{and}\,\,\,\,\gamma(y)=\frac{\mu(y)}{\sigma^2(y)}\,.
\end{equation}

The formula (\ref{runmax}) is a classical expression 
originally derived in \cite{Skorokhod} and used in many later sources including \cite{Olympia}.

\vspace*{5mm}
In the context of the strategy,  the random variable $(M_{T_D(a)}-a)$ corresponds to the final balance of a long position stopped at the trailing stop (with $a$ being the trailing stop threshold), so that the formulae
(\ref{runmax}) and (\ref{Psi}) allow to obtain the distribution of the strategy returns, provided that
the diffusion model is calibrated, functions $\mu(y)$ and $\sigma(y)$ are properly chosen and 
all the parameters are estimated.

\vspace*{1.5mm}
Let us also note that the minimum of the process stopped at a maximum drawup corresponds to a short 
position stopped at the trailing stop. As the cases of long and short position are symmetric,
formulas for their running max/min only differ by the sign of the underlying variable, which allows
us to focus primarily on the long position case (we will call it "D-configuration", which means hitting the lower level, D, first), while for the respective expressions for the short
position case are quite similar and can be obtained as a "mirror reflection", i.e. the change of relevant signs.

\vspace*{0.35cm}
For the case of Ornstein-Uhlenbeck (OU) process, the functions $\mu$ and $\sigma$ are given by
\[ 
\mu(y) = \lambda(\theta - y)\,\,\,\,\,\mbox{and}\,\,\,\,\,\sigma(y)=\sigma\,,
\]
where $\lambda$ is the mean reversion rate, $\theta$ is the (long-term) mean and $\sigma$ the volatility parameter. 

Using these, the function $\Psi$ from (\ref{Psi}) is given by
\begin{equation}\label{PsiOU}
\Psi(u,z)=\exp\left\{\frac{\lambda}{\sigma^2}\left[(z-\theta)^2-(u-\theta)^2\right]\right\}\,.
\end{equation}

Using (\ref{PsiOU}) to calculate the probability of the maximum of OU process stopped at the drawdown
(by the trailing stop) via (\ref{runmax}), numerical integration can be used to obtain the distribution function
and, respectively, the estimate of the probability density function (pdf).

Few examples of the pdf calculated as above for different combinations of parameters, are given on Figure 4.

In each of these examples, the starting point of the process is $1.3$, and the maximum drawdown level is set to be $0.0055$ (which corresponds to the trailing stop of $50$ pips, in the context of the trading strategy).

\vspace*{-0.3cm}
\begin{figure}[ht!]
\centering
\includegraphics[width=90mm]{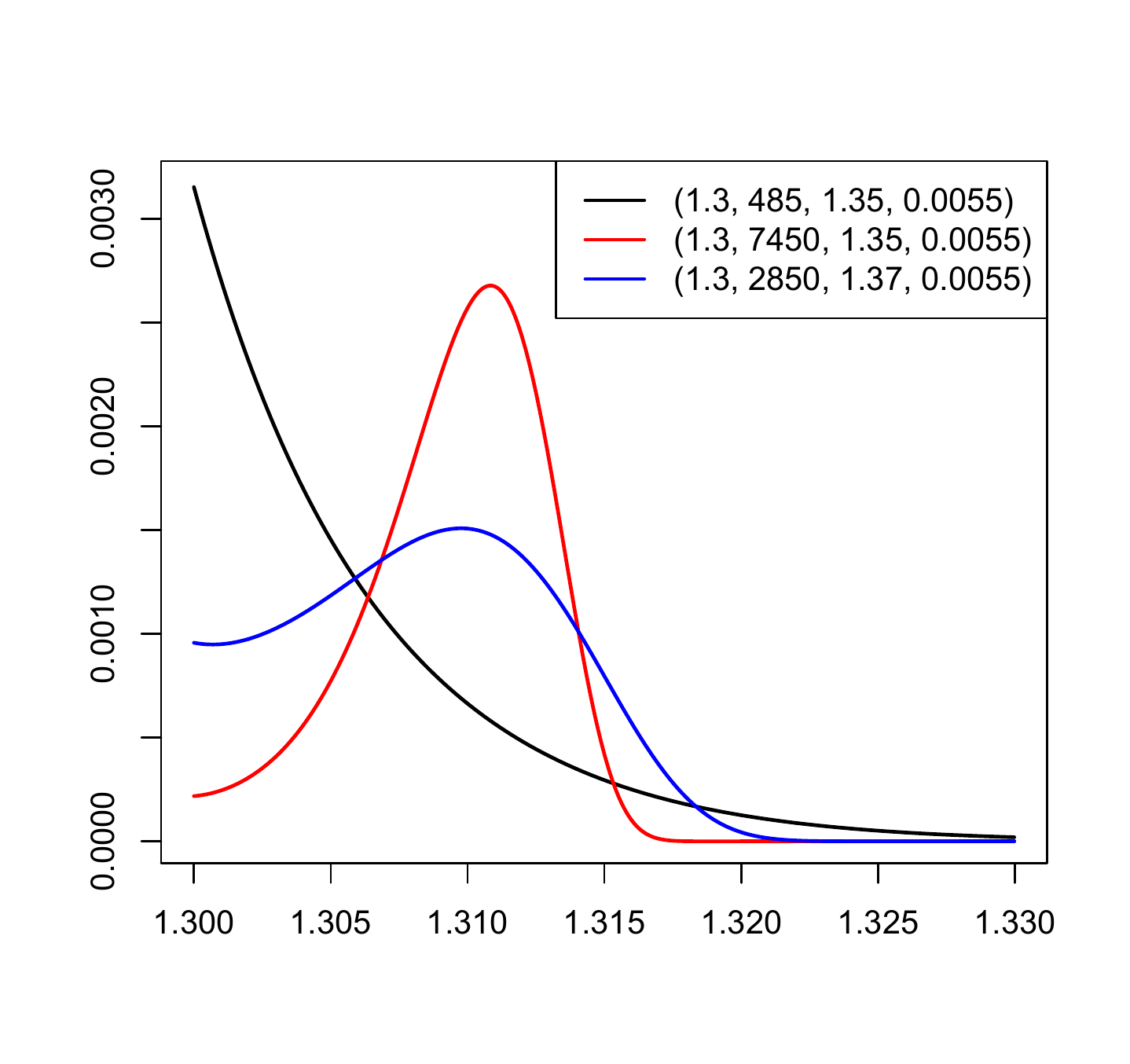}
\vspace*{-0.2cm}
\caption{ Examples of pdf of the maximum process stopped at drawdown. The second parameter
in the set (the one changing from $485$ to $7450$ and down to $2850$)
is the ratio $\lambda/\sigma^2$ and the third parameter is the long-term mean $\theta$.  \label{F3}}
\end{figure}

\vspace*{0.35cm}
Getting back to the point of the strategy realization, recall that, apart from the trailing stop, the 
profit call is also applied to the open position, so that the profile of the weekly strategy return
is actually a composition of profit call probability impact along with the truncated and shifted 
probability density of the running maximum of the process stopped at the drawdown.

\vspace*{0.35cm}
The resulting distribution is a semi-continuous density with a step at point PC, and for the case of OU process it can be calculated explicitly and has a type of profile as illustrated on Figure 5.

\vspace*{-0.3cm}
\begin{figure}[ht!]
\centering
\includegraphics[width=90mm]{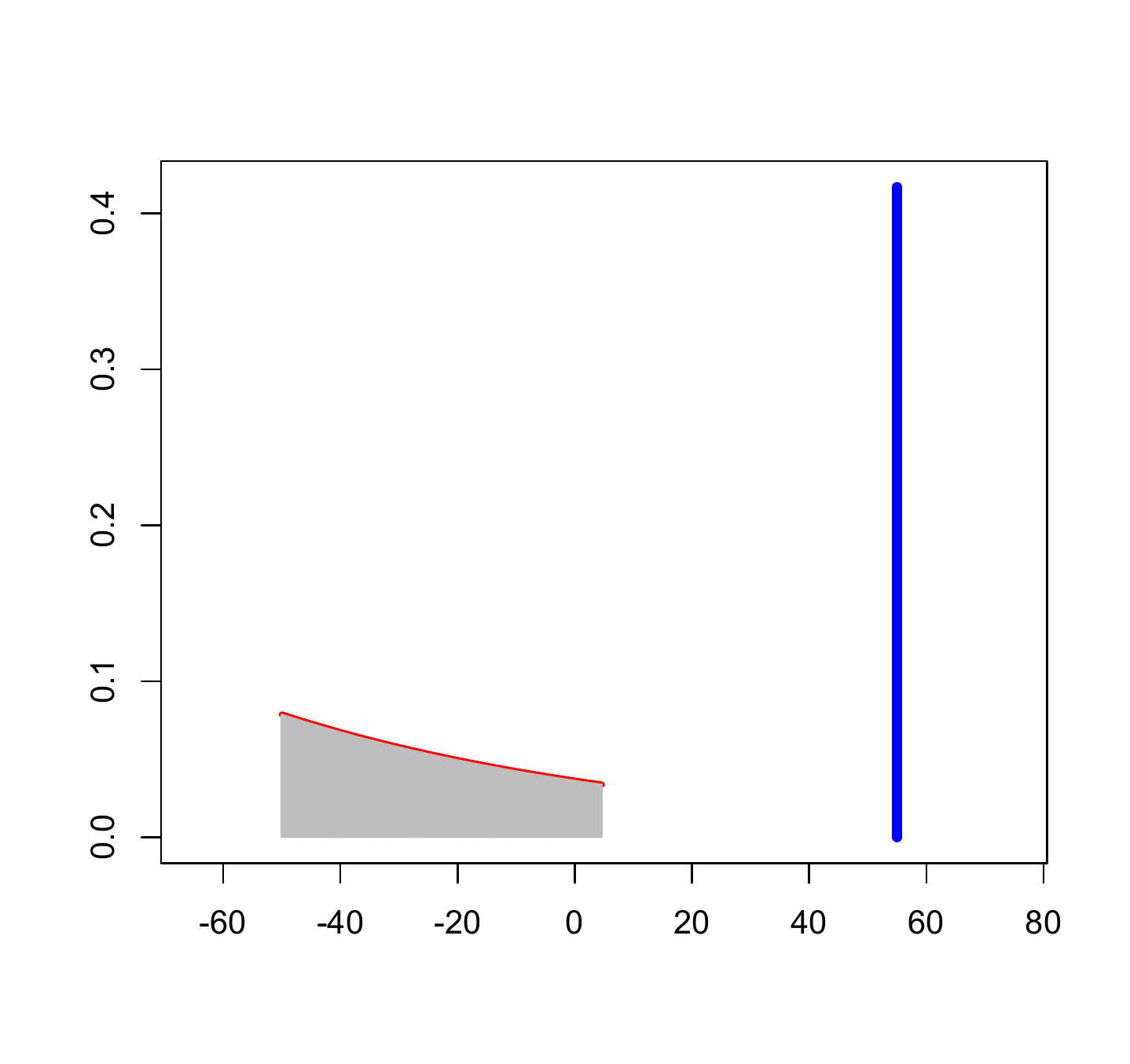}
\vspace*{-0.35cm}
\caption{ Density profile of the weekly returns. The profit call and the 
trailing stop threshold are selected to be, respectively, $55$ and $50$.  \label{F4}}
\end{figure}

\vspace*{0.35cm}
For comparison, the histogram of returns obtained from simulated paths of OU process 
(with parameters similar to the theoretical example above) is given on  Figure 6.

\vspace*{-0.3cm}
\begin{figure}[ht!]
\centering
\includegraphics[width=90mm]{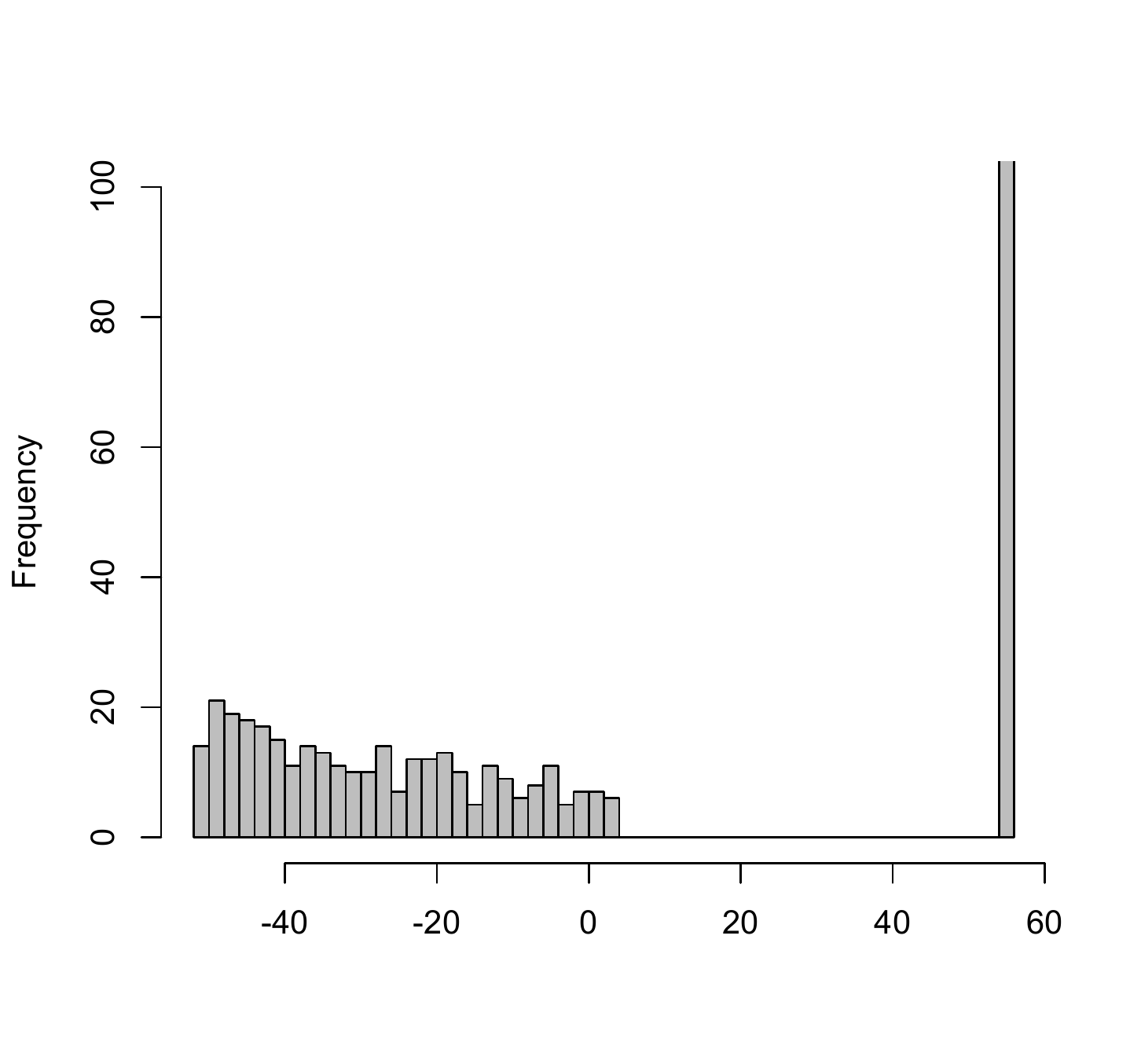}
\vspace*{-0.35cm}
\caption{ Histogram of simulated weekly returns. Parameters are fixed at
same levels as in previous example.  \label{F5}}
\end{figure}

\vspace*{0.35cm}
From the analytic form of pdf as above, required characteristics of the distribution of weekly returns can be estimated.

Some of the most important characteristics are the probability of the profit call and the expected 
value of the weekly returns.
\vspace*{0.4cm}

\vspace*{0.55cm}

Expressed via the absolute price change units (pips), the expected value has the form

\[
\mathbf{E}\left[W_R\right] = \int\limits_{-TS}^{PC-TS}y\cdot f_{M_{TS}-TS}(y)dy + 
PC\cdot \mathbb{P}(M_{TS}\geqslant PC)\,,
\]
where $TS$ and $PC$ are the sizes of the trailing stop and profit call thresholds (expressed in selected units),
$M_{TS}$ is the running maximum of the process stopped at trailing stop and $f_{M_{TS}-TS}$ is the pdf of running maximum shifted down by the value of the trailing stop (effectively, $(M_{TS}-TS)$ is the final balance of the position closed by trailing stop).

%Therefore, the expected value is $PC*$probability plus the expected value of the
%truncated continuous distribution.

\vspace*{0.55cm}
As the size and the value of weekly expected returns is the crucial indicator of the strategy's 
profitability, it might be useful to  have a closer look at borderline cases, such as when expectation changes its sign. 

By fixing the starting point of the process (say, $x_0=1.3$), and the drawdown level ($a=0.005$),
let us trace how expectation changes with the change of the long-term mean, $\theta$
(assume that the parameters $\lambda$ and $\sigma$ are also fixed at that time and the ratio is 
$\lambda/\sigma^2$).

\vspace*{0.35cm}
Table 2 shows the change of expected value with the shift of 
$\theta$, near the edge of the profitability of the strategy (near negative expected value),
considering separately the case when the profit call threshold is the same as the
trailing stop threshold ($PC=TS=0.005$), and when the profit call level is slightly larger $PC=0.0055$ .

\vspace{0.1cm} \vspace{0.1cm}\hspace*{-0.4cm}
\begin{tabular}{|p{1.2cm}|p{1.4cm}|p{1.4cm}|p{1.4cm}|p{1.4cm}|} \hline
& \multicolumn{2}{|c|}{ $PC=0.005$ } & \multicolumn{2}{|c|}{ $PC=0.0055$} \\ \hline
$\theta$ & $\mathbf{P}(PC)$ & $\mathbf{E}\left[W_R\right]$ & 
$\mathbf{P}(PC)$ & $\mathbf{E}\left[W_R\right]$ \\ \hline $1.335$ &
 $0.43$ & $8.4$ &
 $0.42$ & $4.556$ \\ \hline
$1.295$ & $0.36$ & $2.3$ &
 $0.359$ & $\color{red}{-0.77}$ \\
\hline
$1.285$ & $0.34$ & $ 0.72$ &
 $0.34$ & $\color{red}{-2.11}$ \\
\hline
$1.275$ & $0.32$ & $\color{red}{-0.82}$ &
 $0.32$ & $\color{red}{-3.5}$ \\
 \hline
$1.25$ & $0.28$ & $\color{red}{ -4.7}$ &
 $0.28$ & $\color{red}{-6.8}$ \\
 \hline
\end{tabular}

\vspace*{0.15cm}
Table 2.

\vspace*{0.35cm}

Clearly, as the long-term mean drifts away from the process' starting point to the negative side, the probability of hitting PC decreases, as does the mean of the entire profit distribution.

\vspace*{0.35cm}
Arguing slightly ahead of Sections 4 and 5 where we will consider the estimation of underlying parameters
$\theta$, $\lambda$ and $\sigma$ and their implementation in the strategy, we can, at this stage, make the following point: if
the estimation of the parameters from historical data confirm that the 
estimated parameters' current values leads to the positive expected returns, we stick to 
the position opening scheme described in Chapter 2.

If, however, the associated expectation is estimated as negative 
(which, roughly speaking, happens when
the long-term mean is on the "wrong side" of the position opening rate), we might choose not to open the position and skip the week (or wait for the position to hit the opposite threshold opening level -- "U" rather than "D" or "D" rather than "U", ignoring the one that was hit first if it is anticipated as "wrong-way" configuration).

%The full quote (reference) is given in Table in the Appendix.

\vspace*{0.8cm}

\section{Calibration of the OU model}

%Derivation can be found in many sources, such as
%Here, we only quote..
%We quote the least squares regression estimation and maximum likelihood estimation

Estimation of parameters of OU process is well established. Several methods, such Maximum likelihood
and mean squares estimates, can be applied. We only make a brief reference to the estimation
methods here.

It is well known that the explicit solution of the discrete-time version of the OU process (which is
analogous to $AR(1)$ model) is given by
\[
S_{i+1}=S_ie^{-\lambda\delta} + \theta\left(1-e^{-\lambda\delta}\right)+\sigma\sqrt
{\frac{1-e^{-2\lambda\delta}}{2\lambda}}Z\,,
\]
where parameters $\lambda$, $\theta$, are as defined in the previous chapter, $Z$ is the standard Normal random variable and $\delta$
is the discrete-time step size.

\vspace*{0.35cm}
Using the above recursive formula, the Maximum likelihood estimates of the parameters can
be obtained quite straightforwardly (multiple sources can be cited) and they result in

\[ \theta = \frac{S_yS_{xx}-S_xS_{xy}}{n(S_{xx}-S_{xy})-(S_x^2-S_xS_y)}\,, \]

\[ \lambda = -\frac{1}{\delta}\ln \frac{S_{xy}-\mu S_x-\mu S_y+n\mu^2}{S_{xx}-2\theta S_x+n\theta^2}, \]

and

\[ \sigma^2 = \widehat{\sigma}^2\frac{2\lambda}{1-\alpha^2},  \]

where $\widehat{\sigma}^2=\frac{1}{n}[S_{yy}-2\alpha S_{xy}+\alpha^2S_{xx}-2\theta(1-\alpha)
(S_y-\alpha S_x)+n\mu^2(1-\alpha)^2]$, 

with $\alpha = e^{-\lambda\delta}$, and the sums $S_{\circ\circ}$
in the above are as
\[  S_x=\sum\limits_{i=2}^nS_{i-1}\,,\,\,S_y=\sum\limits_{i=1}^nS_{i}
,\,\,S_{xx}=\sum\limits_{i=1}^nS^2_{i-1},\,\,S_{xy}=\sum\limits_{i=1}^nS_{i-1}S_{i}
,\,\,S_{yy}=\sum\limits_{i=1}^nS_{i-1}S_i\,. \]

\vspace*{0.35cm}

Returning to the historical data that we test the strategy with, one of the key questions is:
what history period should be used for estimating the parameters at each time instance, as
the strategy horizon evolves with time?

There is no doubt that the parameters of the underlying OU process keep changing with time, 
but which length of the period is relevant?

%The question is: which history cutoff?..
%There are several ways to estimate the parameters, and the common question is
%-how much of the historical data should be taken into account.

As we have four and a half years, or about $235$ weeks at our disposal, we can 
consider several schemes to approach the dynamic estimation of the parameters, such as:

\begin{itemize}
\item 
At the start of each week of trading, use the data that covers the most recent time period of a fixed length, say, $22$ weeks (about $5$ months).

Use this recent data to estimate the set of parameters of the OU model: 
$\theta$, $\lambda$ and $\sigma$.

By the end of the trading week, the current week's data will be added to the historical dataset, and the data pack to estimate the parameters at the  start of the {\it following week} will include the past week's data plus $21$ previous weeks.

%As we will indicate later, the scheme will be: estimating parameters from %historical data
%(recent 20 weeks) -- and use them to adjust the parameters TS and PC.

%\vspace*{-0.15cm}
%\item 
%As we will indicate later, the scheme will be: estimating parameters from %historical data
%(recent 20 weeks) -- and use them to adjust the parameters TS and PC.

\vspace*{-0.15cm}
\item 
Alternatively, we may use {\bf all} the history accumulated up to the most recent available week, so that the gradually increasing dataset would be used for OU parameters' estimation at the start of each new week.
\end{itemize}

%\vspace*{0.4cm}

\vspace*{0.25cm}
The scheme drafted above is just one possible way for dynamic estimation of parameters.
Certainly, there are alternative ways, such as full Bayesian inference that might be also suitable
in this context (in which case the most recent week could viewed, for example, as {\it newly arrived data} to be used to update the parameters'
estimation and result in {\it posterior distribution} of parameters' vector).

%-Estimate the parameters' set from the range of years of historical data
%$+$ the recent year 

%\vspace*{0.35cm}
%We quote the complete table of estimates of parameters in Table 2.

The impact of the choice of the history period on the estimates of the long-term mean $\theta$ is reflected on the plot in Figure 7. 

\vspace*{-0.3cm}
\begin{figure}[ht!]
\centering
\includegraphics[width=110mm]{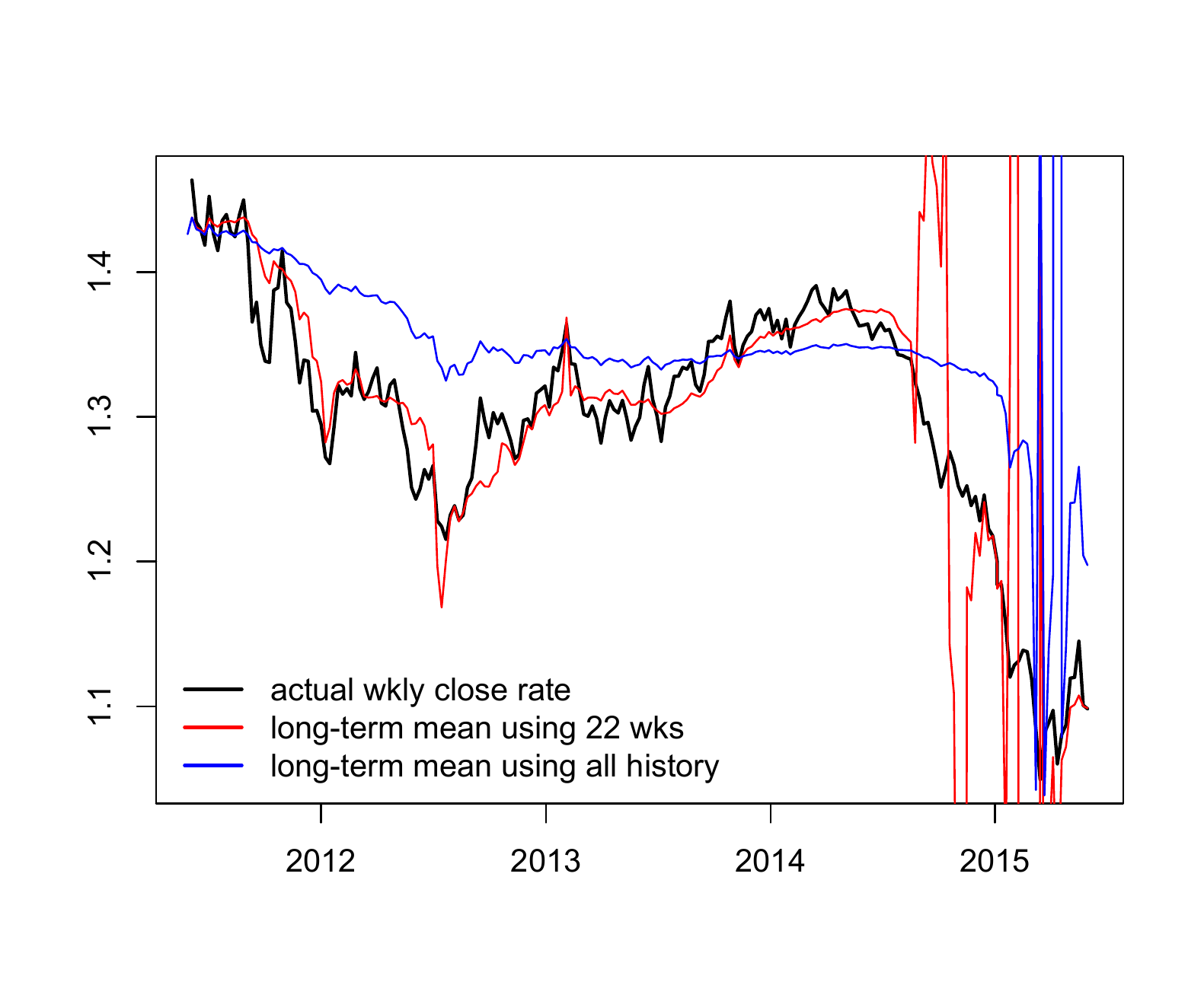}
\vspace*{-0.75cm}
\caption{ Dynamics of the estimates of long-term mean of the underlying OU process,
calculated under different schemes of parameters' updating: 1) using the recent history
(22 recent weeks) only and 2) using all the available history to date -- plotted against
the actual close-of-the-week rate dynamics.  \label{F6}}
\end{figure}

\vspace*{-0.15cm}

As reflected on Figure 7, the most irregular time period is the second half of $2014$ that was marked
by the dramatic fall of the the euro.

The extreme degree of fluctuations of the estimates of $\theta$ (that starts earlier in $2014$ for the recent history updating scheme, and only later towards $2015$ when using the all-history scheme) might be an indication that during that period, the estimates are just not reliable, and the OU process is no longer a good model for the underlying process on such volatile markets.

%Obviously, such periods are not suitable for..

\vspace*{0.4cm}
Next, we also look at the dynamics of parameters' ratio $\lambda/\sigma^2$ -- the rate of convergence 
to the long-term mean over current variance value (as the model distribution depends on this
ratio rather than on each of the parameters alone).

The behavior of the estimates of this ratio, reflected in Figure 8, clearly makes an additional indication of the irregular character of the underlying process towards the second half of $2014$.
% as the model simply fails to fit within the OU process.

\vspace*{-0.15cm}
\begin{figure}[ht!]
\centering
\includegraphics[width=110mm]{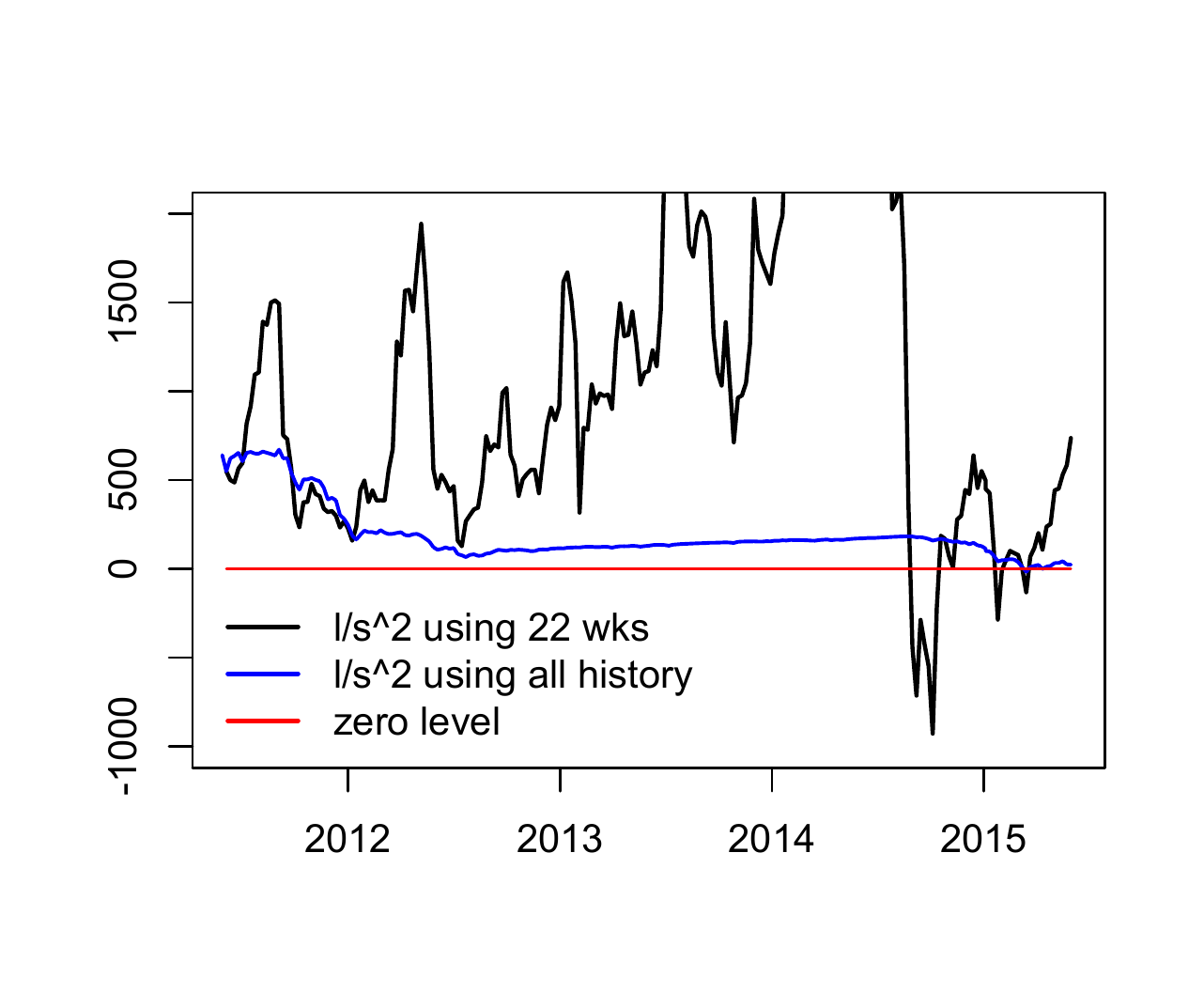}
\vspace*{-0.75cm}
\caption{ Dynamics of the estimates of $\lambda/\sigma^2$ estimated using: 1) the recent history
(22 recent weeks) only and 2) all the available history to date.  \label{F7}}
\end{figure}

\vspace*{0.15cm}
At some point, the rate $\lambda$ of reversion to long-term mean, as estimated from the recent $22$ weeks data, {\it even becomes negative} which means that the model is no longer viable on that time interval.

\vspace*{0.3cm}
Towards the start of $2015$, the estimates seem to converge and the usage of OU might become valid again.
However, the edge of $2014$/$2015$ looks like a clear warning indicating that the models
relying on a particular form of the underlying diffusion process (OU process in this case) should be used with extreme care.

%We quote the estimated of the parameters, referred to selected time %periods, in Table 3 below
%That indicated the time intervals when the parameters' estimates are %reliable ($2011-2013$),
%the interval in which they are clearly not viable ($2014$), followed by the %recent period 
%when they seem to stabilize ($2015$) so that the might process revert to %the OU type yet again.

%As the rate of the reversion decreases,
%On the other hand, the indicator of the long-term mean is clearly telling %us..
During such time intervals, the strategy should probably not be used or only used with additional control measures to exclude the opening that could potentially result in negative returns -- using the indicators such as the ones that we discuss in Chapter 5.

\vspace*{0.15cm}
As already mentioned, the schemes for parameters' updating drafted above are clearly quite simplistic.
%(update + recent year with greater weight)

%or the already mentioned Bayesian inference, where the "fresh" data can %be used to update
%the posterior distribution, while the prior had been estimated using %older data.

\vspace*{0.15cm}
Also,  we do not discuss goodness-of-fit at this point,  and do not provide the justification that the OU is the reliable fit for the data (excluding the second half of $2014$), as opposed to other models.

%of the data on the most time intervals .

Clearly, goodness-of-fit is a separate issue and should
be addressed in more details, which would be beyond the scope of this review paper.

\vspace*{0.07cm}
Here we only note that the strategy outlined in this work, can of course be implemented under different assumptions about the dynamics of the underlying process and different models, such as jump-diffusion processes and related models, for many special cases of which, the weekly returns profile can also be analytically calculated in a way similar to the one we used in Chapter 3.

%Althoug we have done this part of analysis as well, inclusion of this in %the current version 
%of the paper would make its structure too complicated. The issue of %goodness-of-fit 

\vspace*{0.35cm}

\section{Discussion and outlook}

\vspace*{0.2cm}
{\bf Proposed usage of analytical distribution}

\vspace*{0.12cm}
As discussed in Chapter 2, for the first test of the strategy implementation, we use a simple ad-hoc technique to optimize the parameters {\it directly} from historical data. 

\vspace*{0.35cm}
Ideally, of course, the parameters $(U, D ,TS, PC)$ should be estimated from the theoretical distribution of weekly
returns.

%\vspace*{0.05cm}
%Ideally, we would estimate the parameters from the task of maximizing %the expected profit
%-and then use the above updating scheme (update + recent year with %greater weight)

%Alternatively, we These parameters
%are used for the next week.. and so on.. the system gets updated.

A proper scheme for updating the $(U, D ,TS, PC)$ set could look like:
\vspace*{0.1cm}
\begin{itemize}
\item
Estimate the OU parameters in a way such as outlined in Chapter 4.
\item
Use the OU parameters estimates to calculate the analytical distribution of
weekly returns, as described in Chapter 3, for a range of the strategy parameters $(U, D ,TS, PC)$
% remaining unknown  in a way such as outlined in Chapter 4.
\item
Pre-set the desirable level of probability of a Profit call (say, $PC>=40\%$) and/or the desired expected value of weekly returns,
and optimize the set $(U, D ,TS, PC)$ such that the analytical predicted
values of the PC probability and weekly returns would fit into the desired levels.
%from the {\it analytical }
%Use this to calculate TS and PC
\end{itemize}

\vspace*{0.05cm}
That way, we would be solving a sort of the inverse problem - find $(U, D ,TS, PC)$ such that optimize the expected profit.

%\vspace*{0.05cm}
%The problem is that the long term mean is difficult to estimate with %required confidence.

Of course, this optimization problem, despite being quite straightforward in its description, is actually a problem of quite a significant computational complexity.

\vspace*{0.25cm}
However, even if we stick to the simple optimization scheme described in Chapter 2 at this stage, we could still use similar ideas, to help identify the FALSE position opening cases (the ones with negative predicted expected returns and with relative small probability of PC) and exclude them.

That would increase the efficiency of the strategy even under the simple scheme for parameters' updating and optimization.

\vspace*{0.25cm}
In order to make the first rough check of whether the predicted expected value of weekly returns and probability of a Profit Call are in correspondence with actual results for the returns of the strategy implemented under the simple optimization scheme, we can just compare these indicators directly.

%\vspace*{0.35cm}
%\begin{itemize}
%\item
%Alternatively, use it not to cut off the undesirable (false) positions %opening
%\end{itemize}

\vspace*{0.15cm}
We select a time interval from the history -- the year $2013$, -- and
 plot actual returns according to the test run of the strategy as in Chapter 2 (at this point we use first of the schemes for the parameters' optimization
described in Chapter 2 -- with optimization w.r.t the last year history only and also first of the OU parameters' updating schemes from Chapter 4 -- using the $22$ recent weeks only) against the expected values estimated via the OU estimates.
The results are given in Figure 9.

\vspace*{-0.15cm}
\begin{figure}[ht!]
\centering
\includegraphics[width=110mm]{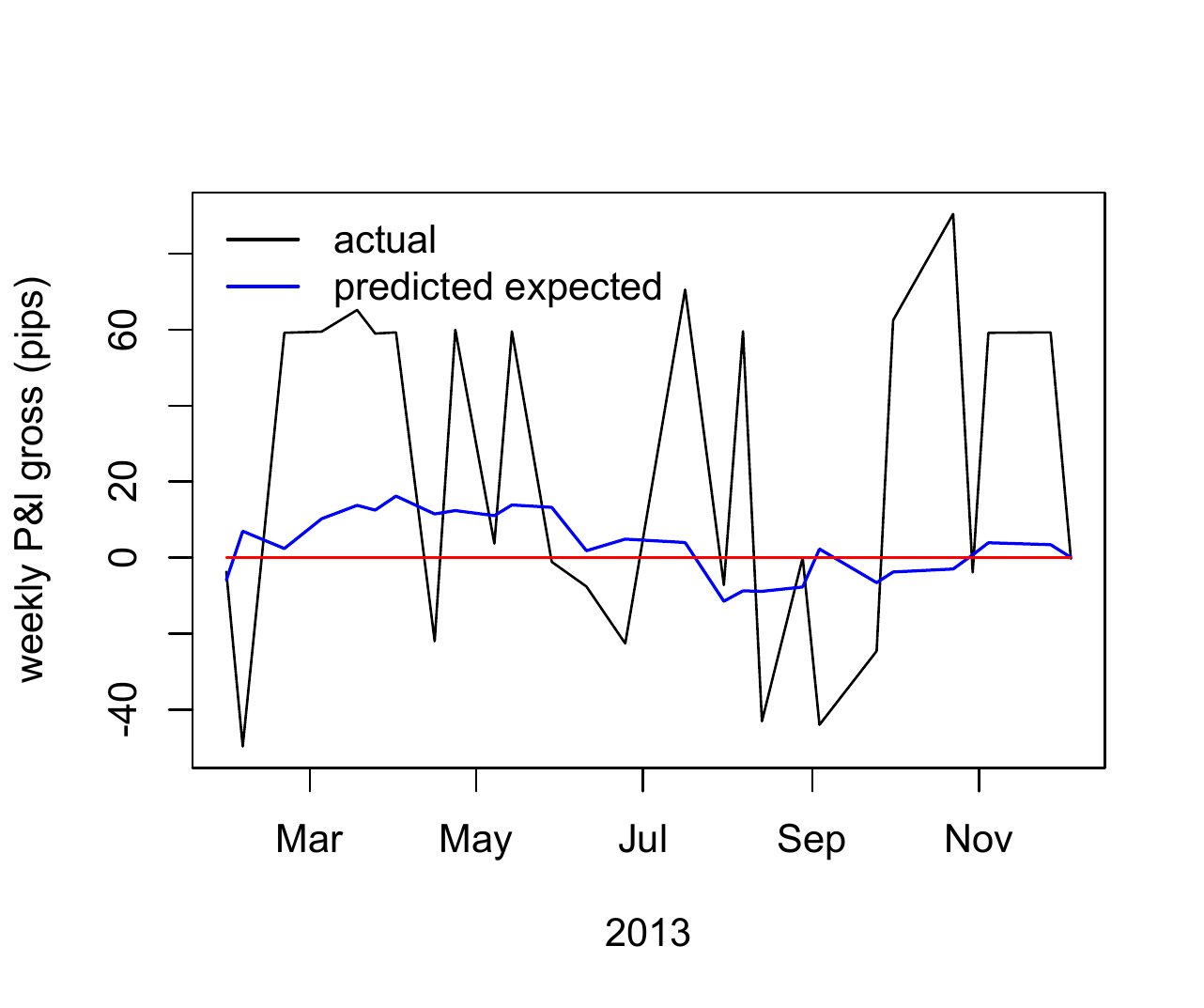}
\vspace*{-0.55cm}
\caption{ Dynamics of actual weekly returns against the calculated expected value for
a selected time interval.  \label{F8}}
\end{figure}

\vspace*{0.2cm}
{\small For simplicity, we only consider the "D" configuration -- when the price hits "D" threshold
within the week first, and the long position opens. 
Combining "D" and "U" in one figure would complicate the visual analysis by mixing upward- with downward trends.
Of course, the "U" configuration can be analyzed in the same way.}

\vspace*{0.2cm}
As observed in Figure 9, there is a certain correlation between expected (predicted) values and actual returns (although this can could only be used to get a general impression, and a more accurate analysis would be needed for practical implementation).

\vspace*{0.3cm}
%Specifically, for each week we note the level $D$ and calculate expected %returns using 
%{\it pre-set parameters} $TS, PC, ...$

%\vspace*{0.15cm}

%\vspace*{0.8cm}
%Another illustration of the possible use of this technique, revisiting.. %Following Figure 8 and Table 3.
%As the rate of the reversion decreases, the probability of hitting PC %decreases too, as does 
%the mean of the entire profit distribution. In Table 4 below.

Some further visual analysis (though again quite a loose one, but this may be viewed as initial approach) for the whole available history is indicated in Figure 10.
Again, we compare the expected values (calculated at start of each week) with actual return by the end
of corresponding week
(again, we only look at those open positions that correspond to "D" configuration).

\vspace*{-0.15cm}
\begin{figure}[ht!]
\centering
\includegraphics[width=110mm]{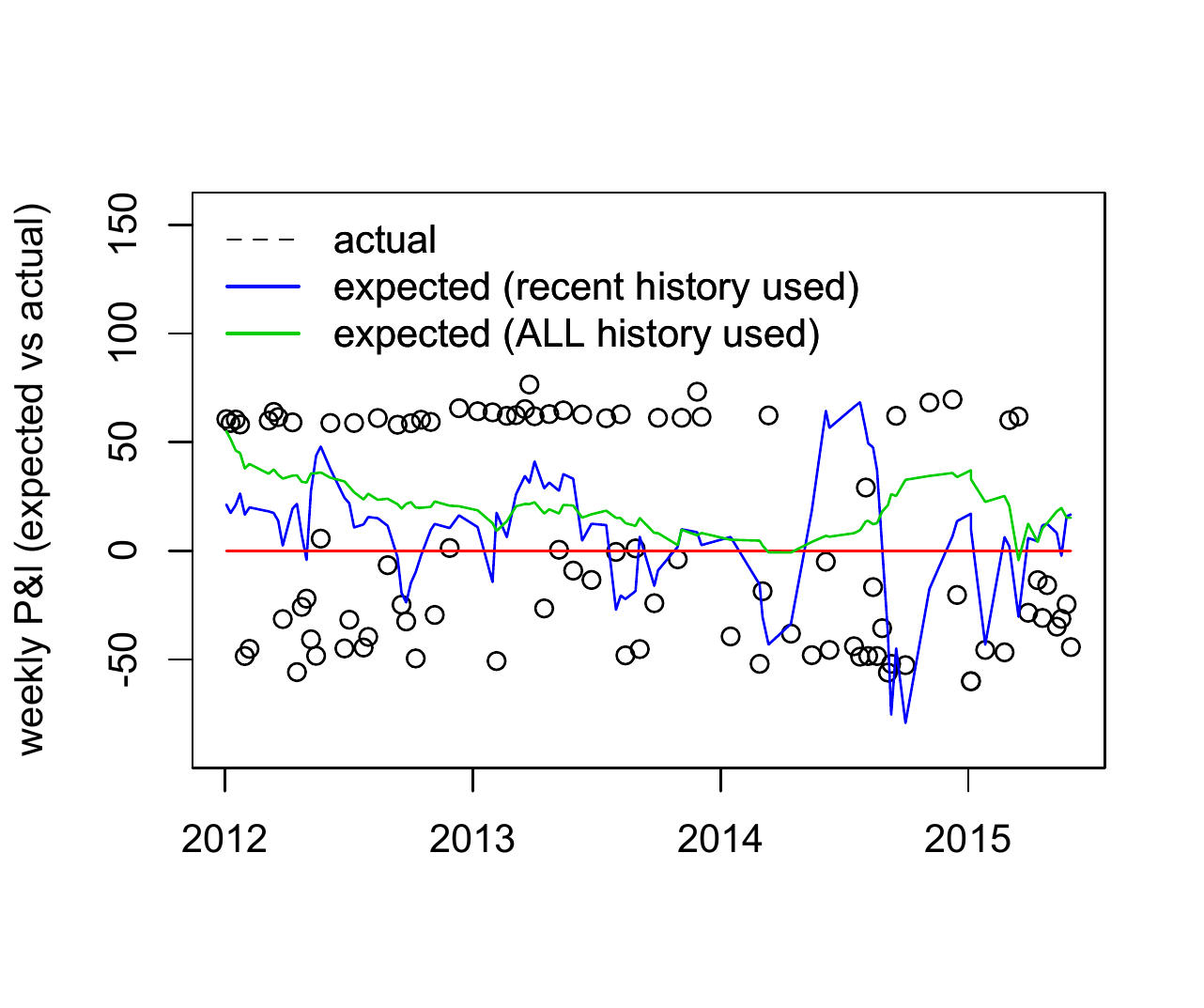}
\vspace*{-0.55cm}
\caption{ Dynamics of actual weekly returns against the calculated expected value for
the entire time interval.  \label{F9}}
\end{figure}

\vspace*{0.2cm}
Here, we use both of the schemes for parameters updating as described in Chapter 4: the first one uses the recent $22$ weeks of data only, the other uses all the history available by the moment of parameters' estimation (on the 
weekly basis).

As initial visual analysis shows, before $2014$, both schemes seem to be in good correspondence with
each other, and also there appears to be certain match with the tendency of the actual returns.

From $2014$, however, the discrepancy between the mean values obtained using two schemes,
becomes very large. The estimate that comes from scheme that uses the whole history, at least,
seems to predict the significant decline in the expected values around the first half of $2014$,
which in reality would be a considerable warning sign for a user of the strategy, whereas the estimate
coming from the parameters' updating scheme that uses $22$ recent weeks of data only, appears to completely fail predicting the  behavior of actual returns throughout the $2014$ crisis. 

Towards the start of $2015$, both schemes seem to converge again and revert to a better prediction 
of actual returns dynamics (although the "prediction" still seems to be questionable in $2015$,
probably due to the continuing volatile and rough behavior of the market).

%\vspace*{0.3cm}
%{\bf General observations}

\vspace*{0.35cm}
{\bf Comparison of the calculated probability of the Profit Call with its actual frequency }

\vspace*{0.2cm}

Finally, we look at the behavior of such crucial indicator as the probability of the process hitting the
 Profit Call threshold.

This probability can be estimated using analytical formula for the maximum of the
diffusion process (provided that the PC level is hit before the TS threshold) as described 
in Chapter 3.

If we introduce now a simple Bernoulli r.v. that takes value $1$ in case of the Profit Call and $0$ otherwise
then we can consider the sequence of such r.v.'s, each of which corresponds to each next week of
trading (certainly, the probability of the PC, say, $p_i$, keeps varying from week $i$ to the next one) and if we sum up all these Bernoulli r.v.'s over a time period, the resulting sum will be a {\it Poisson Binomial r.v.}  with expected value $\sum\limits_0^n p_i$ (the sum of success probabilities) and variance $\sum\limits_0^n(1-p_i) p_i$.

\vspace*{0.1cm}
As each of probabilities $p_i$ is pre-calculated at start of corresponding week, we can therefore
calculate the resulting (theoretical) expected value of this Poisson Binomial r.v. (effectively, this would
be an "average theoretical probability of a profit call" over the period) and we can compare
it with the {\it actual} frequency of weeks with profit calls (simply dividing the number of
weeks in which PC was hit, by the total number of weeks in the period).

The results are given in Table 3 (again we consider only the "D" configuration here for simplicity and
consistency).

\vspace{0.1cm} \vspace{0.1cm}\hspace*{-0.4cm}
\begin{tabular}{|p{3.2cm}|p{1.2cm}|p{3.1cm}|p{2.4cm}|} \hline
$\,$  & actual & \multicolumn{2}{|c|}{theoretical}  \\ \hline
 $\,$ & $\,$  & with $22$ wks history & with all history\\ \hline 
 mean & $0.388$ &
 $0.354$ & $0.381$ \\ \hline
 variance &$0.240$ & $0.225$ & $ 0.236$  \\
 \hline
\end{tabular}

\vspace*{0.2cm}
Table 3. Relative number of actual profit calls vs. theoretical values

\vspace*{0.3cm}
Again, the PC estimate that comes from the scheme of parameters' updating that
uses all the history available by the corresponding week, is much closer to the 
observed relative number of the actual profit calls, with respective variances (of
the corresponding Poisson Binomial r.v.'s) also in good agreement.

\vspace*{0.8cm}

%To summarize, we will be looking to take advantage on one of the %following price movement schemes:

%--Ud   (Up-before-drawdown)
%--Du   (down-before-drawup)

\section*{Acknowledgments}
We acknowledge and appreciate the collaboration with the Research and Development team of Quantum Brains Capital who provided feedback that allowed for useful and motivating discussions.


\vspace*{0.8cm}

\begin{thebibliography}{500}

\bibitem{Skorokhod}
Gihman, I. I. and A. V. Skorokhod (1972).
\newblock{Stochastic Differential Equations, Springe-Verlag, New York.}

\bibitem{Olympia}
L. Pospisil, J. Vecer and O. Hadjiliadis (2009).
\newblock{Formulas for stopped diffusion processes with stopping times based on drawdowns and drawups.}
\newblock{ Stochastic Processes and their Applications, 119 (8), 2563--2578.}


\end{thebibliography}
\end{document}